# Observation of pseudogap-like feature above $T_c$ in LiFeAs and $(Ba_{0.6}K_{0.4})Fe_2As_2$ by ultrafast optical measurement

Kung-Hsuan Lin,[1,*] Kuan-Jen Wang,[1] Chung-Chieh Chang,[1] Yu-Chieh Wen,[1] Dzung-Han Tsai,[1] Yu-Ruei Wu,[1] Yao-Tsung Hsieh,[1] Ming-Jye Wang,[1] Bing Lv,[2] Paul Ching-Wu Chu,[2] and Maw-Kuen Wu[1,3,+]

[1] Institute of Physics, Academia Sinica, Taipei 11529, Taiwan

[2] Texas Center for Superconductivity, University of Houston, Houston, TX 77004, USA

[3] Department of Physics, National Dong Hwa University, Hualien 97401, Taiwan

**Abstract**

We utilize ultrafast optical measurement to study the quasiparticle relaxation in stoichiometric LiFeAs and nearly optimally doped $(BaK)Fe_2As_2$ crystals. According to our temperature-dependent studies of LiFeAs, we have observed pseudogap-like feature at onset temperature of ~ 55 K, which is above $T_c$ = 15 K. In addition, the onset temperature of pseudogap ~90K was also observed in $Ba_{0.6}K_{0.4}Fe_2As_2$ ($T_c$ = 36 K). Our findings seem implying that the pseudogap feature, which is due to antiferromagnetic fluctuations, is universal for the largely studied 11, 111, 122, and 1111 iron-based superconductors.

[*] linkh@phys.sinica.edu.tw

[+] mkwu@phys.sinica.edu.tw

## 1. Introduction

The recently discovered iron-based high-temperature superconductors have attracted great attention.[1] It was reported that $(Ba,K)Fe_2As_2$ (122 system) and CeFeAs(O,F) (1111 system) exhibit a rich phase diagram with antiferromagnetism (or spin density wave, SDW) at low doping and superconducting (SC) state at intermediate doping without structural or phase transition.[2,3] On the other hand, LiFeAs (111 system) and FeSe (11 system) were found to be superconducting without additional doping.[4,5] Since the phase diagram appears very different, it raises a question if there is universal property among the largely studied 11, 111, 122, and 1111 iron-based superconductors. Is its SC mechanism similar to that of the cuprate superconductors?

One of the general features in cuprate superconductors is the existence of pseudogap above $T_c$.[6] The low-energy pseudogap in cuprate superconductors is believed to be the precursor of the SC gap. The high-energy pseudogap in cuprate superconductors was not found in overdoped regime.[6] However, the reports on the observation of pseudogap in iron-based superconductors are only handful among the huge amount of research papers. Recently, Kasahara *et al.* clearly showed the existence of pseudogap was on $BaFe_2(AsP)_2$, and the pseudogap existed from underdoped to overdoped regime until the SC state disappeared.[7] The mechanism of the pseudogap was proposed to resulting from orbital ordering, which was related to the electronic nematic transition. Moreover, pseudogap was also observed in overdoped regime in Na(FeCo)As.[8] The experimental evidence so far has shown that the presence of pseudogap in iron-based superconductors (in overdoped regime) is different from that in cuprate superconductors. Pseudogap in iron-based superconductors was also observed by angle-resolved photoemission spectroscopy (ARPES),[9-11] scanning tunneling microscopy (STM),[8] resistivity measurement,[12] infrared optical measurement,[13,14] and ultrafast optical measurement.[15-20] Ultrafast optical time-resolved measurement is not surface-sensitive, and is a useful tool to study the bulk property of the materials. There have been several studies reporting ultrafast phenomena in iron-based superconductors.[15-26] By observing the temperature-dependent quasiparticle-relaxation, pseudogap-like feature have been observed in FeSe (11),[18] underdoped (BaK)Fe$_2$As$_2$ (122),[16] Ba(Fe,Co)$_2$As$_2$ (122),[17] Ca(Fe,Co)$_2$As$_2$ (122),[20] and SmFeAsO$_{0.8}$F$_{0.2}$ (1111).[15] However, the pseudogap feature of LiFeAs, which is the representative crystal in 111 systems, has not been reported yet. LiFeAs is special among the iron-based superconductors. First, the structure is the simplest in FeAs-based superconductors. Second, it does not have static antiferromagnetic order while most of (1111), (122) and (111) parent compounds have. Third, it is superconducting in stochastic compound. Fourth, it does not have structural transition while FeSe has. Studies of LiFeAs might be helpful to understand if there exists universal property among iron-based superconductors.

In this work, we have utilized ultrafast optical measurement to study quasiparticle dynamics in LiFeAs and nearly optimally doped (BaK)Fe$_2$As$_2$ (BKFA) single crystals. For LiFeAs, we found the onset temperature of pseudogap feature is ~ 55 K, which is above the SC temperature $T_c$ ~ 15 K. According to our fitting results, the gap $\Delta_{PG}$ is on the order of 13 meV. We argue that the pseudogap of LiFeAs results from antiferromagnetic fluctuations and is not associated with the SC gap. However, this does not imply that the SC paring mechanism is not associated with antiferromagnetic instability. Moreover, the SC gap $\Delta(0) \approx 5\text{meV}$ of LiFeAs was obtained with the assumption of BCS gap-temperature relation. On the other hand, the SC gap $\Delta(0) \approx 12\text{meV}$ of nearly optimally doped BKFA was also obtained. Similar to previous results of underdoped BKFA,[16] the quasiparticle relaxation behavior, which appeared in SC state, persisted up to ~ 90 K. This suggests that the pseudogap occurs above $T_c$ (36 K) in nearly optimally doped BKFA, and is the precursor of superconductivity.

## 2. Experimental Details

The magnetic susceptibilities of LiFeAs and BKFA as a function of temperature were measured as shown in Figs. 1 (a) and (b). The SC temperatures have been defined as 15 K and 36 K, respectively. For optical measurement, both single crystals were cleaved to reveal a shining surface, and mounted on the holder of cryostat in Ar-filled glove box. After the cryostat was moved out from glove box, the pressure of the chamber was immediately lowered to below $10^{-4}$ mtorr for avoiding oxidation.

A typical nondegenerate pump-probe measurement was conducted. 800 nm probe pulses and frequency-doubled 400 nm pump pulses were used from an 80MHz Ti:sapphire oscillator. In order to minimize laser heating effects, the repetition rate was reduced to 8 MHz with a pulse picker. The pump was modulated at ~ 1MHz with an acousto-optical modulator (AOM) and the optical fluence was 5~10 μ J/cm$^2$. The full width of half maximum (FWHM) of the temporal cross-correlation of pump and probe pulse was ~500 fs. The relatively longer duration was due to the dispersion of the 400 nm pump pulses through the 5cm-thick AOM crystal. A color filter was placed in front of the photodetector for eliminating leakage pump light. We recorded the reflectivity of the probe pulse as a function of time delay. Typically, signals with changes on the order of $10^{-6}$ could be resolved with our experimental setup.

## 3. Experimental Results and Analysis

Figure 2 shows a few representative traces of time-resolved reflectivity changes of LiFeAs at different temperatures. Overall, the traces reveal similar features above 99 K as shown in Fig. 2 (a). A fast relaxation component with negative magnitude appears at ~ 0 ps, and the relaxation time is comparable to the FWHM (0.5 ps) of our temporal cross-correlation of pump and probe pulses. After the ultrafast relaxation, there is a negative hump centered at ~ 7 ps, followed by a relatively slow relaxation component. In Fig. 2 (b), new additional feature before 5 ps appears below 51 K. The magnitude of this relaxation component increases with decreasing temperature.

First of all, the ultrafast relaxation time within 0.5 ps at all temperatures should be attributed to thermalization of non-Fermi distribution of electrons namely electron-electron scattering.[27] In our experimental condition that the optical pump fluence kept the same, the magnitudes of this ultrafast relaxation did not show significant dependence to temperature. Secondly, the negative humps at ~ 7 ps, followed by a slow relaxation, also appear at all temperatures. We ascribe this feature to the effects of propagating coherent longitudinal acoustic phonon and quasiparticle diffusion along the depth direction. In transparent or semi-transparent media, the propagation of coherent acoustic phonons can induce temporal sinusoidal oscillation, with multiple cycles, due to coherent Brillouin scattering.[28] The oscillation period can be resolved and is determined by the refractive index, sound velocity, optical wavelength, and incidence angle of the optical probe in the media. However, the feature of time-resolved optical reflectivity due to propagating phonon in highly absorptive materials, such as the samples we studied, is not trivial. In order to understand the signals due to propagating phonons, we have used finite difference time domain (FDTD) method[29] to simulate the time evolution of strain distributions in LiFeAs. Since the complex refractive index and photoelastic constant of LiFeAs were unknown, we used the optical parameters of FeSe to qualitatively understand this feature.

The time-resolved optical reflectivity can be represented as[30]

$$\Delta R(t) = \int_0^\infty f(z) S_{33}(z,t) dz, \tag{1}$$

where

$$f(z) \propto \left[ \frac{\partial n}{\partial S_{33}} \sin\left(\frac{4\pi n z}{\lambda} - \phi\right) + \frac{\partial \kappa}{\partial S_{33}} \cos\left(\frac{4\pi n z}{\lambda} - \phi\right) \right] e^{-z/l}. \tag{2}$$

$S_{33}$ is the longitudinal strain, $n$ and $\kappa$ are real and imaginary part of the complex refractive index, $z$ is the depth position from the surface, $\lambda$ is the optical wavelength, and $l$ is the optical absorption length. $\phi$ is a constant, which is related to $n$ and $\kappa$.[30] $\partial n / \partial S_{33}$ and $\partial \kappa / \partial S_{33}$ are real part and imaginary part of the photoelastic constant, respectively. Following Eq. (1), Fig. 3 (a) shows the calculated $\Delta R(t)$ by using photoelastic constants with different phases. The centers of the negative hump vary for different phases of photoelastic constants. Therefore, one could not determine the phonon oscillation period simply from the first dip time. Although the unknown optical and photoelastic constants made quantitative analysis of LiFeAs difficult, the negative humps observed in Fig. 2 for all temperatures could still be well explained by the calculated $\Delta R(t)$ due to propagating phonons in Fig. 3 (a). Note that we did not consider the diffusion of quasiparticles and incoherent phonons (heat) in our calculation. The slow relaxation after the hump could be due to the quasiparticle/heat diffusing out from the optical probe region near the surface. Diffusion of quasiparticles probably dominates the relaxation within 40 ps because quasiparticles usually diffuse faster than heat does with the same initial distribution.

In Fig. 2 (b), the additional relaxation components before 5 ps are attributed to pseudogap-like feature. Similar phenomena were also observed in FeSe,[18] SmFeAsO$_{0.8}$F$_{0.2}$,[15,19] and Ba(Fe,Co)As$_2$.[17] In order to extract this relaxation component for quantitative analysis, it requires subtraction of the phonon signals. The black line in Fig. 3 (b) represents the averaged traces of 130 K, 135 K, 140 K, 145 K, and 150 K. After smoothing by adjacent averaging, the red line was used for removing the phonon signals at all temperatures. Fig. 4 (a) shows the processed results for 20 K, 40K and 51 K. We used two exponential time decay functions and one step function to fit the processed traces:

$$A(t) = A_e e^{-t/\tau_e} + A_p e^{-t/\tau_p} + A_0, \qquad (3)$$

where the first and second exponential terms account for electron thermalization and quasiparticle relaxation due to pseudogap, respectively. $A(t)$ was convoluted with the cross-correlation of pump and probe pulses to fit the time-resolved reflectivity changes $\Delta R(t)/R$.

Fig. 4 (b) shows the fitted $A_p$ as a function of temperature, and the corresponding time constant $\tau_p$ are also shown in the inset. Below 55K, $A_p$ begins to be non-zero and the corresponding $\tau_p$ are in the range of 1.5 ps and 2 ps. The increase of reflectivity component $A_p$ at low temperature might be trivially ascribed to lower electronic and lattice specific heats. But we found the magnitude of $A_p$ roughly saturated below 20K, and thus ruled out this explanation. Instead, we used a bottleneck condition of quasiparticles in a gapped system to explain the observed phenomenon. The change of photoinduced reflectance is generally assumed proportional to the quasiparticle population,[15,16,23] and is proportional to the quasiparticle density in the excited state of a gapped system in a low excitation limit.

The bottleneck condition can be described by[15]

$$\frac{\Delta R_{PG}}{R} \propto \left[1 + B \exp\left(-\frac{\Delta_{PG}}{k_B T}\right)\right]^{-1}, \qquad (2)$$

where $\Delta R_{PG}$ is the pseudogap-induced change of photoreflectance and $\Delta_{PG}$ is the effective pseudogap magnitude. $B = 2\nu / [N(E)\hbar\Omega_c]$, where $\nu$ is the number of bosons involved in the relaxation process across the pseudogap, $N(E)$ is the density of states at Fermi surface, and $\Omega_c$ is the cutoff frequency of the bosonic spectrum. To obtain the effective pseudogap magnitude, we treated $B$ as a fitting factor. The red line in Fig. 4 (b) shows the fitting curve and we have obtained $\Delta_{PG} = 13 \pm 1$ meV.

For temperature around and below $T_c$ (15K) of LiFeAs, we reduced the repetition rate of the laser pulses to further minimize the laser heating effects. We found negative step appeared below 15K. We have also done fluence-dependent studies, and the negative step was nonlinear and disappeared at high fluence. We ascribe this component to quasiparticle relaxation due to SC gap. After the pump pulse excites quasiparticles near the surface of LiFeAs, the transient temperature exceeds $T_c$. High frequency bosons with energy $\omega \geq 2\Delta$ are created in the excitation volume and break the Copper pairs. The relaxation time reflects the population of high frequency bosons or the recovery time of SC state when the heat escapes from the optically probed depth. In order to extract the SC component, the traces below 15 K were subtracted by the trace at 16 K (, which is just above $T_c$). Fig. 5 (a) shows the subtracted traces at 7 K, 10 K and 13 K. The relaxation of the quasiparticles is far longer than our observation window within 50 ps and appears like a step function. In the case of Ba(Fe,Co)$_2$As$_2$, the SC recovery time can be up to nanosecond time scale.[17]

The magnitude of the reflectivity change $A_S$, due to the SC quasiparticle relaxation, was obtained by averaging the subtracted traces after 20 ps. Fig. 5 (c) shows the magnitude of $A_S$ as a function of temperatures. We used Rothwarf-Taylor (RT) model to fit the data.[16] The density of thermally-excited quasiparticles, $n_T(T)$ is proportional to $\left[|A_S(T)|/|A_S(T \to 0)|\right]^{-1} - 1$, and $n_T \propto \sqrt{\Delta(T)T} \exp(-\Delta(T)/k_B T)$. We assumed that $\Delta(T)$ follows BCS temperature dependence with $\Delta(0)$ as a fitting parameter. The fitting lines of $n_T(T)$ and $|A_S(T)|$ are shown in Figs. 5 (b) and (c) for $\Delta(0)$=3.5 meV, 5.2 meV, and 6.0 meV. Although the SC gap could not accurately determined from our fitting results, the range of $2\Delta(0)$ between 7 meV and 12 meV still agrees with previous reports.[31-34]

We have also studied temperature-dependent relaxation of quasiparticles in nearly optimally doped BKFA crystal. Fig. 6 shows the time-resolved optical reflectivity of probe pulses as a function of time delay at four representative temperatures. At 110 K as shown in Fig. 6 (d), the trace shows a fast exponential function with time constant of < 1 ps, followed by a feature due to propagating acoustic phonons as discussed previously. Similarly, the fast relaxation is ascribed to electron thermalization. Below $T_c$ (36 K) of BKFA, an additional slow exponential component with time constant of several tens of picoseconds appears as shown in Figs. 6 (a) and (b). The slow relaxation is attributed to the recovery time of SC state.

We have subtracted the contribution of acoustic phonon signals and fit the quasiparticle relaxation as a function of temperature. The red line in Fig. 6 (d), which was the smoothed trace at 110K by adjacent averaging, was used for subtraction. Similar to Eq. (3) for LiFeAs, we have used $A_e e^{-t/\tau_e}$ to account for electron thermalization, and $A_S e^{-t/\tau_S}$ to account for the SC state recovery. Figs. 7 (a) and (b) show the temperature-dependent $\tau_S$ and $A_S$. The magnitude of SC term $A_S$ dramatically increases below $T_c$ (36 K) and roughly saturates below 25 K. The red line in Fig. 7 (b) shows the fitting curve by using the RT model. The fitted SC gap $\Delta(0) = 12$ meV $\pm 1$ meV agrees with previous reports.[11,35-37]

## 4. Discussions

The origin of pseudogap is still controversial even for relatively long studied cuprate superconductors. It is also interesting whether the behavior of pseudogap in iron-based superconductor is similar. The pseudogap of underdoped BKFA, with SDW transition, has been observed with optical conductivity measurement,[38] ARPES,[9] and ultrafast optical measurement.[16] The onset temperature T* of pseudogap for underdoped BKFA is above $T_c$, which is similar to that of underdoped cuprate superconductors. It was suggested that antiferromagnetic fluctuations drive both the pseudogap and superconductivity in BKFA, and the pseudogap is possibly the precursor of superconductivity.[9,16,38]

The presence of pseudogap in optimally doped BKFA is somewhat controversial.[11,35] From Fig. 7 (b), we have noticed that the magnitude of $A_S$ does not vanish above $T_c$, indicating the pseudogap behavior up to ~ 90 K. This phenomenon was also observed in underdoped BKFA.[16] However, Chia *et al.* also conducted ultrafast optical measurement on optimally doped BKFA and did not observe significant tail behavior above $T_c$.[16] We argue that our studied $(Ba_{0.6}K_{0.4})Fe_2As_2$ should be slightly underdoped since our $T_c$ = 36 K is slightly lower than the $T_c$ = 37 K of BKFA reported in Ref. 16. We found similar situation for ARPES measurement that pseudogap was found in BKFA with $T_c$ = 35 K[11] but was not clearly found in BKFA with $T_c$ =37 K.[35] In addition, the pseudogap of slightly underdoped BKFA was also found by optical conductivity measurement.[13] Kwon *et al.*

found that the pseudogap at 100 K continuously evolved to SC gap below $T_c$,[13] which also agrees with our observation that the SC relaxation component persists up to ~ 90 K. Compared with our results and previous studies, pseudogap indeed appears in slightly underdoped BKFA, which does not have SDW transition. And our experimental results also support that the pseudogap in BKFA should be the precursor of the SC gap. Up to now, no evidence of pseudogap has been found in overdoped BKFA, which is similar to high-energy pseudogap in cuprate superconductors. However, this behavior is different from Co-doped and P-doped $BaFe_2As_2$, that the pseudogap was found from underdoped regime to overdoped regime and vanished with superconductivity.[7,10,14] In addition, pseudogap was also found in overdoped $Ca(FeCo)_2As_2$.[20] It was reported that the infrared pseudogap in Co-doped and P-doped $BaFe_2As_2$ is unrelated to superconductivity,[14] which is also different from the argument that the pseudogap is the precursor of superconductivity in BKFA. Note that K atoms are doped out of the FeAs planes while Co and P are doped in the FeAs planes. It was suggested that in-plane doping or out-of plane doping would affect how SC gap form.[37] Similar situation may occur to the formation of pseudogap in (Ba122) or (Ca122) systems.

Different from nearly optimally doped BKFA, we have found two different relaxation components for LiFeAs, indicating that the pseudogap and SC gap are weakly coupled. Similar to in-plane-doped (Ba122),[14] 1111 system,[19,39] and $Na(Fe_{1-x}Co_x)As$,[40,41] our experimental results reveal that the pseudogap feature of LiFeAs should not be associated with the SC gap. The pseudogap in P-doped or Co-doped (Ba122)[14] and 1111 system[19,39] was reported to associate with antiferromagnetic fluctuations. Unlike 122 and 1111 system, LiFeAs is superconducting without dopants and static SDW transitions. However, previous reports did indicate that SDW fluctuation was observed in SC state and normal state either by nuclear magnetic resonance[42,43] or neutron scattering techniques.[44,45] The onset temperature of pseudogap feature of LiFeAs we studied was found to be ~ 55 K, which also coincides with the observation that antiferromagnetic fluctuations already exist above 40K in normal state.[42] For other (111) materials, the spin gap of 13 meV was found in non-superconducting $Li_{0.94}FeAs$ crystal with no static antiferromagnetic order.[46] Moreover, the pseudogap feature was also found in $Na(Fe_{1-x}Co_x)As$ in underdoped and overdoped regimes.[40] The properties of pseudogap in $Na(Fe_{1-x}Co_x)As$, including gap size, shape, and T-evolution with onset temperature of 54 K are similar to that of SDW gap in parent NaFeAs.[40] These evidences lead us to suggest that the pseudogap in LiFeAs should also result from antiferromagnetic fluctuations. Note that Zhou *et al.* found the pseudogap in $Na(Fe_{1-x}Co_x)As$ is a local phenomenon without long-range ordering, and ruled out the simple band-structure effect as a possible origin.[40] The pseudogaps we mentioned in LiFeAs and that in $SmFeAs(OF)$[15,19] are also not a simple band-structure effect. For example, the pseudogap of LiFeAs $2\Delta_{PG} \approx 26$ meV does not mean a 26 meV gap at some arbitrary point in the Brillouin zone.

Despite the different answers to whether the pseudogap is the precursor of superconductivity in BKFA and other FeAs superconductors, it is still quite universal that antiferromagnetic fluctuations are believed to formation of pseudogap in FeAs superconductors[9,14-16,19,38-41] and in cuprate superconductors.[6] Moreover, the pseudogap in FeSe (11 system) was also reported to be from some sort of short-range order, and magnetic fluctuations cannot be excluded.[18] The main question would be what the driving force to antiferromagnetic fluctuations is. This open question is out of the scope of this report. However, the structure of LiFeAs is the simplest among FeAs superconductors. LiFeAs is also superconducting in stochastic compounds without SDW transitions. In addition, LiFeAs does not have structural transition while FeSe does. Our observation of pseudogap in LiFeAs could be helpful to unveil the mechanism of antiferromagnetic fluctuations, which has been still widely believed to be associated with superconducting gap and pseudogap in iron-based and in cuprate superconductors.

Finally, we briefly discussed the SC gap energy of LiFeAs and BKFA obtained from our experimental results. According to ARPES measurement results of optimally doped BKFA, $\Delta$ is ~ 12 meV at inner/outer electron pockets ($\gamma/\delta$ bands) and inner hole pocket ($\alpha$ band) while $\Delta$ is ~ 6 meV at outer hole pocket ($\beta$ band).[11,35,36] Due to the multiband feature of iron-based superconductors, the time-resolved optical reflectivity should be governed by the contribution of quasiparticle relaxation in all bands. Our obtained $\Delta(0) = 12 \pm 1$ meV agrees well with that of $\alpha, \gamma, \delta$ bands in BKFA measured by ARPES. As for LiFeAs, $\Delta$ of $\alpha, \beta, \gamma, \delta$ bands, measured by ARPES, are 5.0, 2.5, 4.2, 2.8 meV, respectively.[34] Our fitted $\Delta(0) \approx 3.5-6$ meV of LiFeAs, as shown in Fig. 5 (c), also agrees well with that of $\alpha, \gamma$ bands, measured by ARPES. However, we found that our fitting model seems being dominated by the larger SC gaps among all bands. In contrast, the optical and thermodynamic measurements pick the smallest gap size in the entire Brillouin zone.[13,31,33]. Although the quasiparticle dynamics at different bands could be resolved with different relaxation times,[23] we did not observe another significant SC quasiparticle relaxation components. One of possible explanations would be the quasiparticle relaxation of bands with smaller gaps are too fast and within our pulse duration. Another possibility would be relaxation times of all bands are similar and could not be resolved. This might explain we got better fitting for BKFA, as shown in Fig. 7 (b) since $\Delta \approx 12$ meV for three bands among four bands in BKFA. But we got relatively worse fitting $\Delta(0) \approx 3.5 - 6$ meV for LiFeAs since $\Delta$ ranges from 2.5 – 5 meV among four bands in LiFeAs.

## 5. Conclusions

We have studied LiFeAs and nearly optimally doped BKFA with ultrafast optical techniques. The pseudogap feature was found in LiFeAs with onset temperature of ~ 55 K. The formation of pseudogap in LiFeAs is not associated with the SC gap. On the other hand, the pseudogap feature was also found in nearly optimally doped BKFA with onset temperature of ~ 90 K. Our experimental results suggest this pseudogap could be the precursor of SC gap. It seems the pseudogap feature is universal among the widely studied 11, 111, 122, and 1111 iron-based superconductors, and is related to antiferromagnetic fluctuations. Our observation of pseudogap in LiFeAs, without doping and structural/magnetic transition, could be helpful to explain the driving force of antiferromagnetic fluctuations, which has been still widely believed to be associated with superconducting gap and pseudogap in iron-based and cuprate superconductors.

## 6. Acknowledgements

The authors would like to thank T.-M. Chuang and Y. Matsuda for stimulating discussions. This work is sponsored by National Science Council of Taiwan under Grant Nos.: NSC100-2112-M-001-028-MY3.

**Figure Captions**

**Figure 1** The temperature-dependent dc magnetic susceptibility of (a) LiFeAs and (b) $(Ba_{0.6}K_{0.4})Fe_2As_2$ measured in field cooling (FC) and zero field cooling (ZFC) modes. The insets show the magnetic susceptibility around superconducting transition temperature.

**Figure 2** The time-resolved optical reflectivity of LiFeAs at temperatures (a) above 99K and (b) below 99K. The traces above 99K are similar while that below 99K have temperature-dependent relaxation component before 5 ps.

**Figure 3** (a) The simulated time-resolved optical reflectivity due to propagating acoustic phonons in highly absorptive material. The position of dips varies with different phases of photoelastic constant. Note that the traces are arbitrarily scaled for easier comparison. (b) The black line represents the averaged trace of LiFeAs at 130K-150K. The red line represents the further smoothed trace by adjacent averaging for subtraction of signals due to phonon propagation and quasiparticle diffusion in the depth direction.

**Figure 4** (a) The processed time-resolved optical reflectivity without phonon signals at 20K, 40K, and 51K. (b) The temperature-dependence of experimental (in black dots) and fitted (in red line) magnitudes of $A_p$ in Eq. (3), due to the pseudogap quasiparticle relaxation in LiFeAs. The temperature-dependence of the corresponding relaxation time $\tau_p$ are shown in the inset.

**Figure 5** (a) The processed time-resolved optical reflectivity at 7 K, 10 K, and 13 K,

that subtract the trace at 16K (just above $T_c$ = 15 K of LiFeAs) to reveal the SC quasiparticle relaxation. (b) The density of thermally-excited quasiparticles, $n_T(T)$ and (c) the magnitude of $A_S$ described in the text are shown in black dots. The fitting lines with $\Delta(0)$=3.5, 5.2, and 6.0 meV are also shown.

**Figure 6** The time-resolved optical reflectivity of nearly optimally doped BKFA at (a) 14K, (b) 33K, (c) 49K, and (d) 110K. The red curve, obtained by smoothing the trace at 110K with adjacent averaging, is used for subtraction of signals due to acoustic phonons and other effects such as quasiparticle diffusion.

**Figure 7** The dots represent (a) $\tau_S$ and (b) $|A_S|$ of SC quasiparticle relaxation term in nearly optimally doped BKFA as a function of temperature. The red line represents the fitting curve with $\Delta(0)$=12 meV.

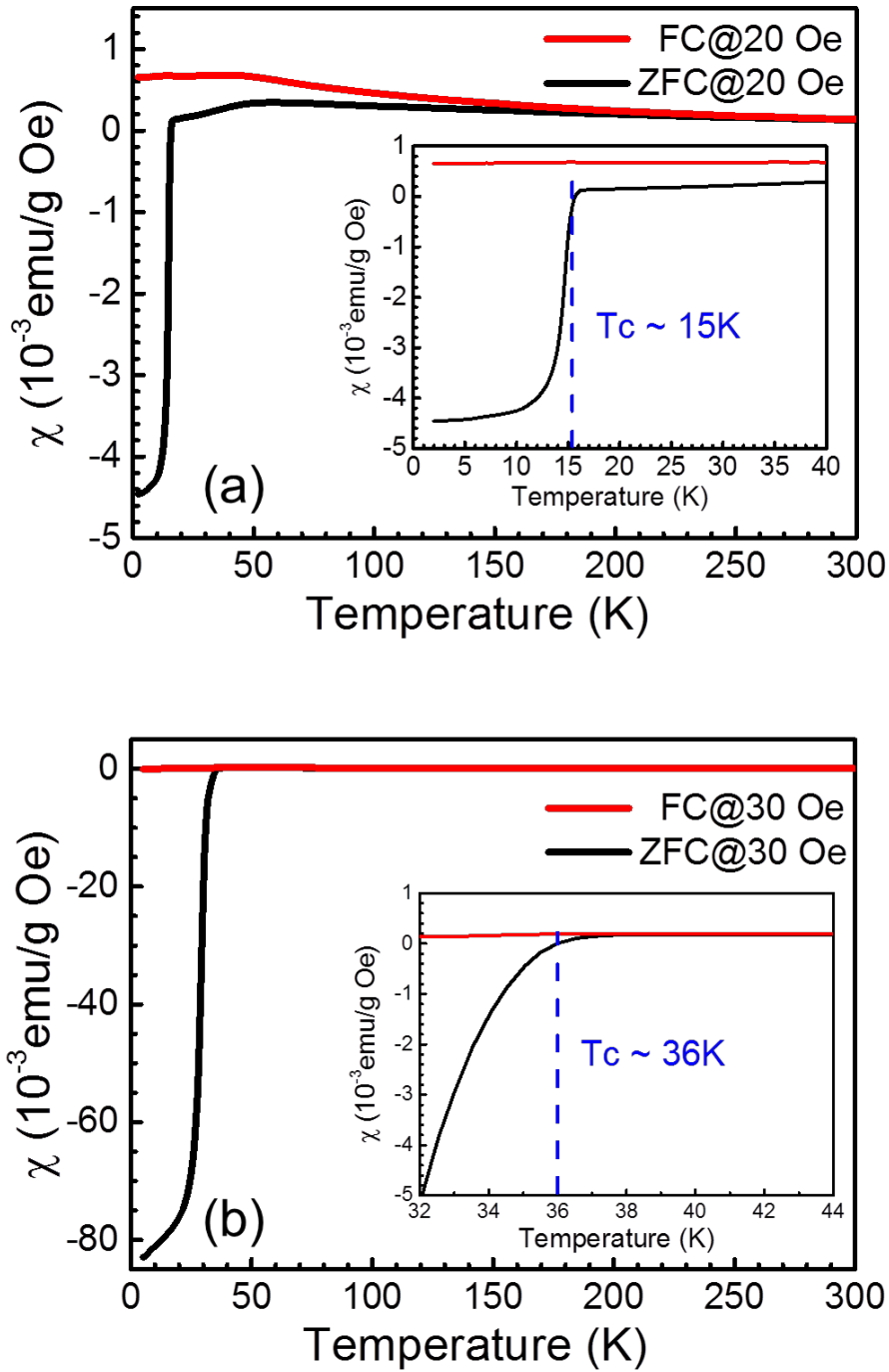

**Figure 1** The temperature-dependent dc magnetic susceptibility of (a) LiFeAs and (b) $(Ba_{0.6}K_{0.4})Fe_2As_2$ measured in field cooling (FC) and zero field cooling (ZFC) modes. The insets show the magnetic susceptibility around superconducting transition temperature.

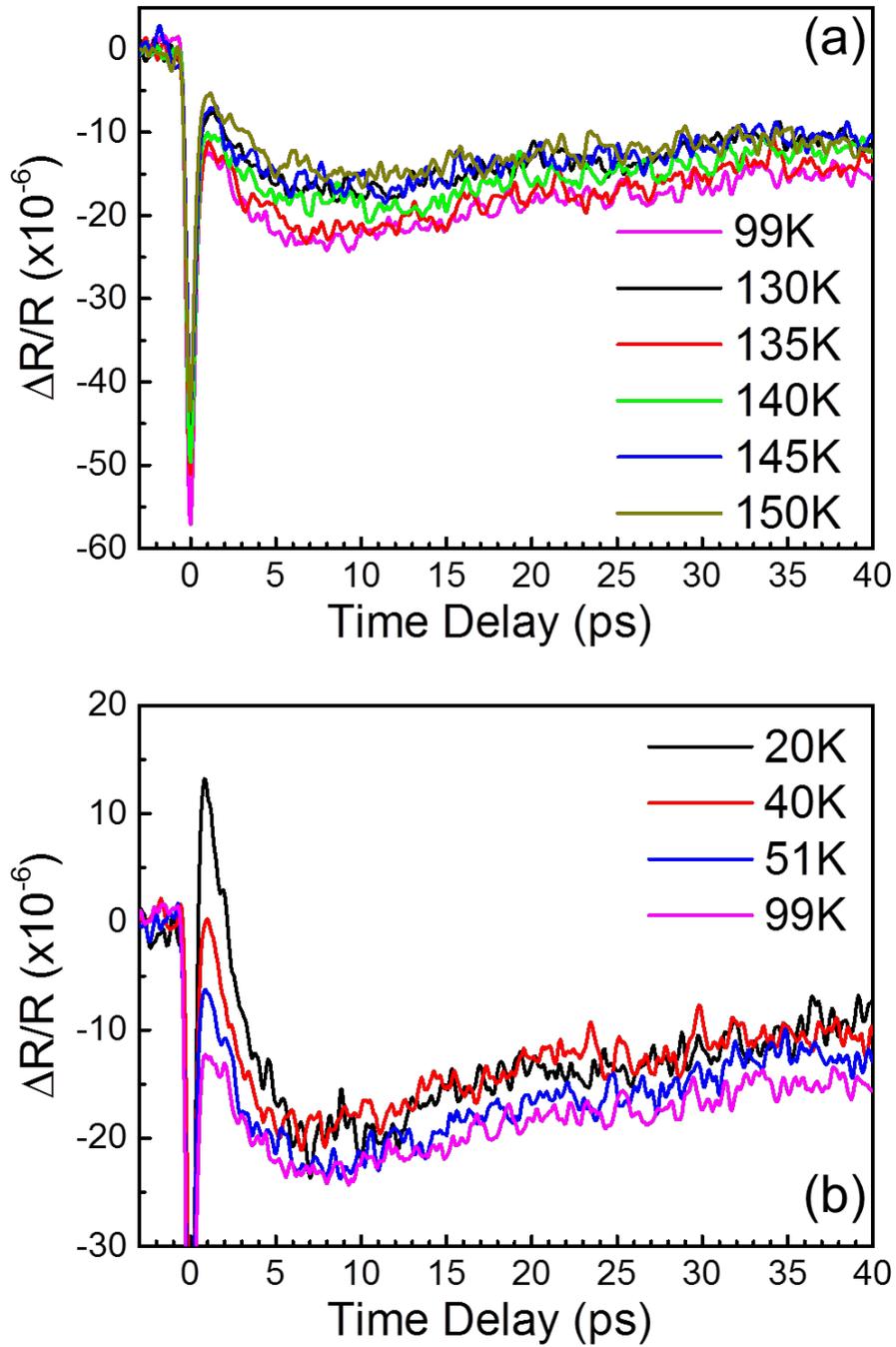

**Figure 2** The time-resolved optical reflectivity of LiFeAs at temperatures (a) above 99K and (b) below 99K. The traces above 99K are similar while that below 99K have temperature-dependent relaxation component before 5 ps.

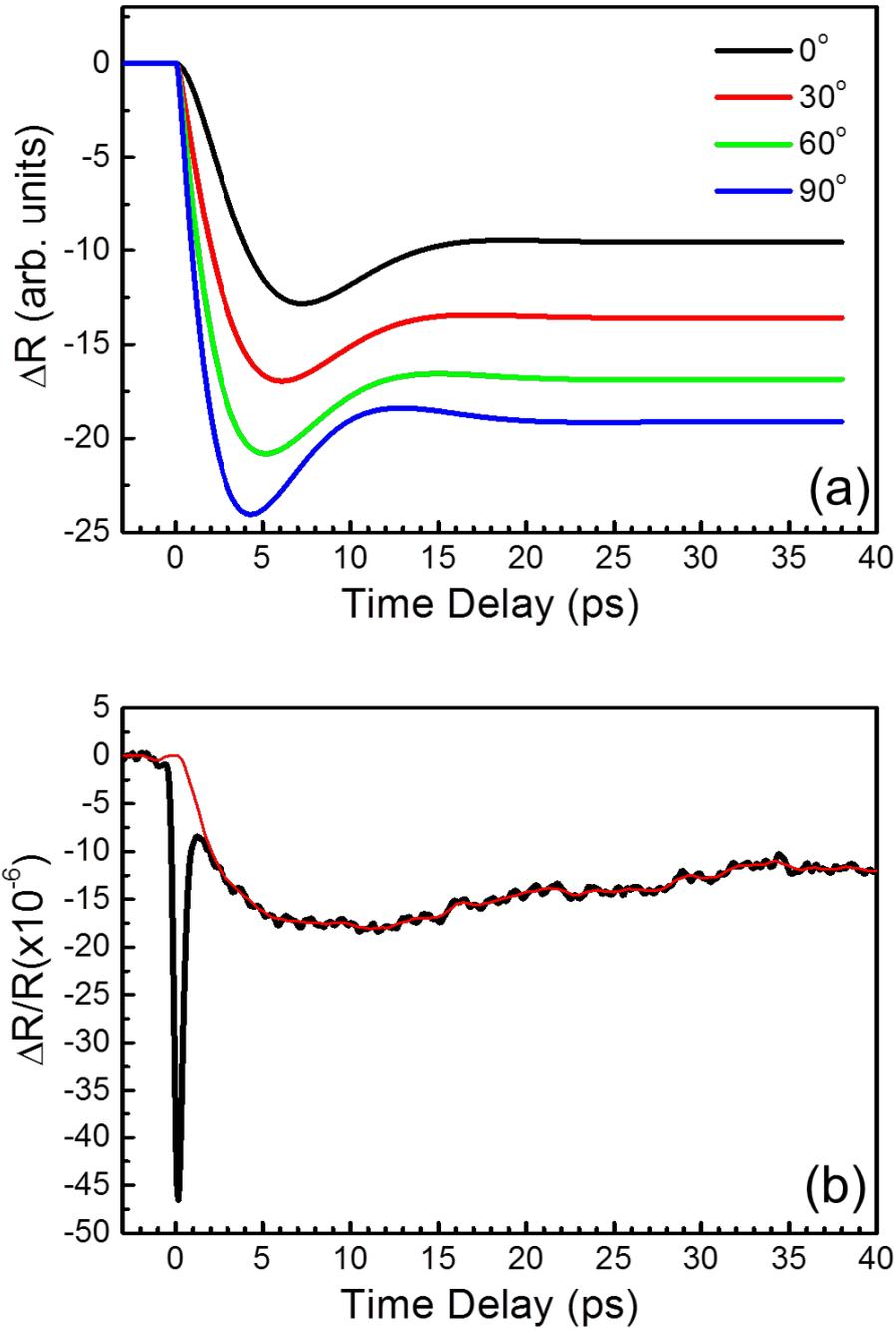

**Figure 3** (a) The simulated time-resolved optical reflectivity due to propagating acoustic phonons in highly absorptive material. The position of dips varies with different phases of photoelastic constant. Note that the traces are arbitrarily scaled for easier comparison. (b) The black line represents the averaged trace of LiFeAs at 130K-150K. The red line represents the further smoothed trace by adjacent averaging for subtraction of signals due to phonon propagation and quasiparticle diffusion in the depth direction.

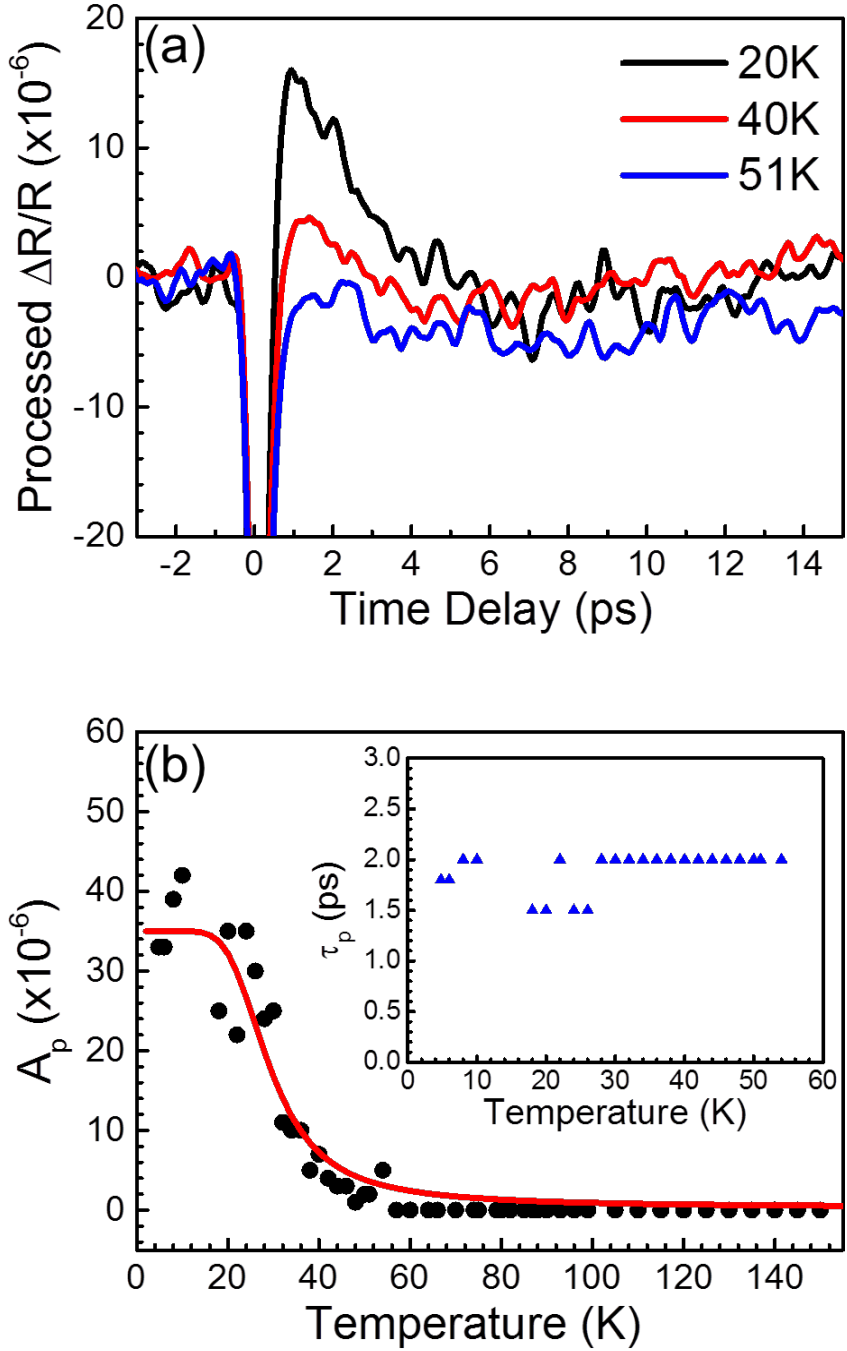

**Figure 4** (a) The processed time-resolved optical reflectivity without phonon signals at 20K, 40K, and 51K. (b) The temperature-dependence of experimental (in black dots) and fitted (in red line) magnitudes of $A_p$ in Eq. (3), due to the pseudogap quasiparticle relaxation in LiFeAs. The temperature-dependence of the corresponding relaxation time $\tau_p$ are shown in the inset.

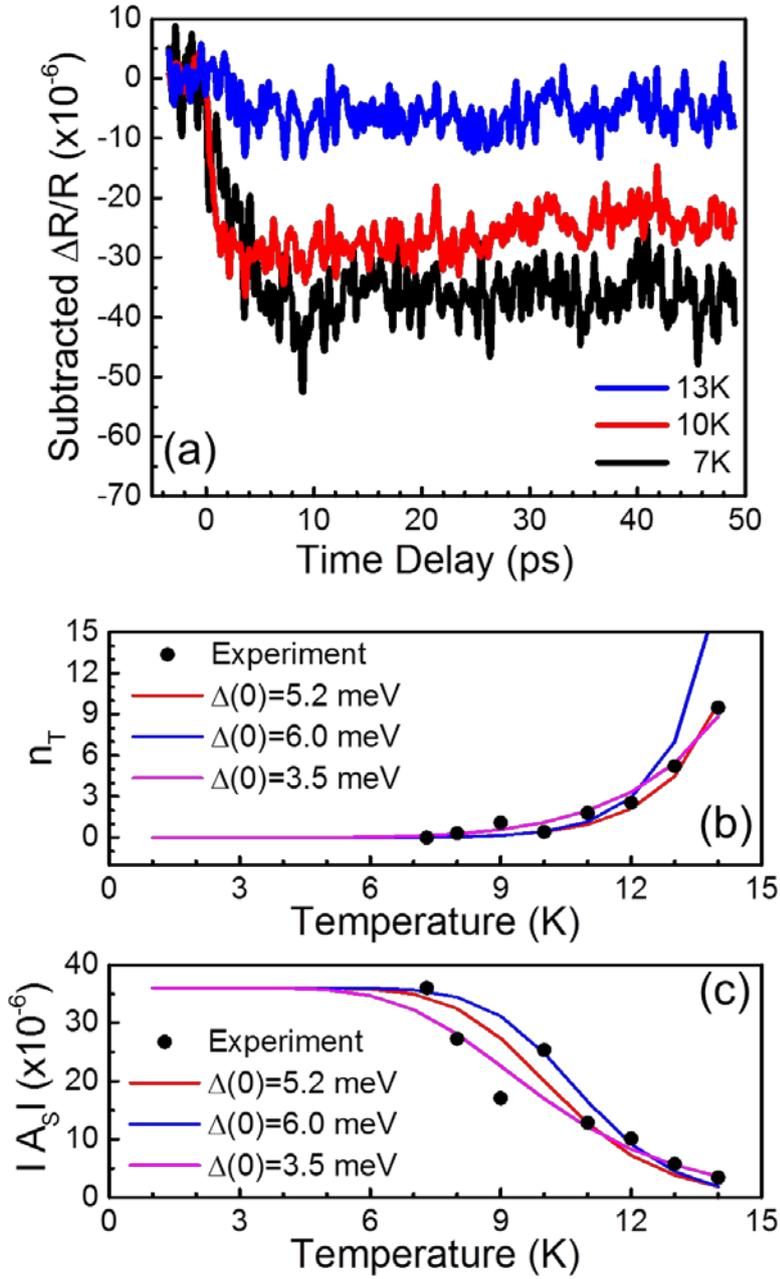

**Figure 5** (a) The processed time-resolved optical reflectivity at 7 K, 10 K, and 13 K, that subtract the trace at 16K (just above $T_c$ = 15 K of LiFeAs) to reveal the SC quasiparticle relaxation. (b) The density of thermally-excited quasiparticles, $n_T(T)$ and (c) the magnitude of $A_S$ described in the text are shown in black dots. The fitting lines with $\Delta(0)$=3.5, 5.2, and 6.0 meV are also shown.

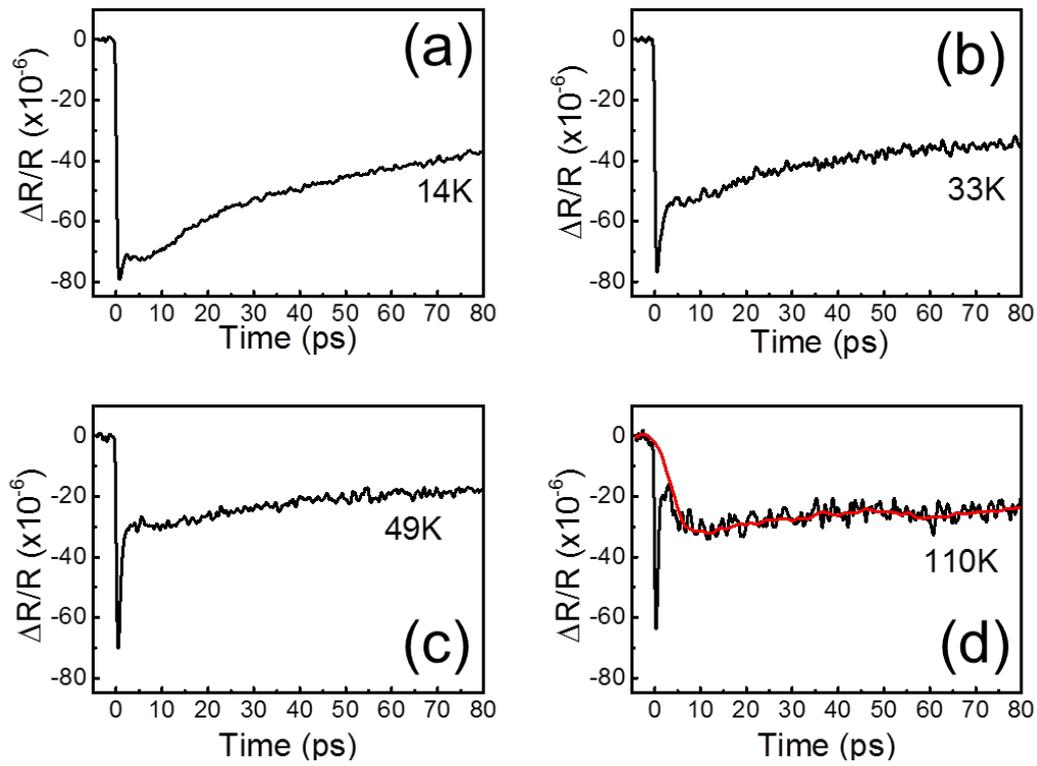

**Figure 6** The time-resolved optical reflectivity of nearly optimally doped BKFA at (a) 14K, (b) 33K, (c) 49K, and (d) 110K. The red curve, obtained by smoothing the trace at 110K with adjacent averaging, is used for subtraction of signals due to acoustic phonons and other effects such as quasiparticle diffusion.

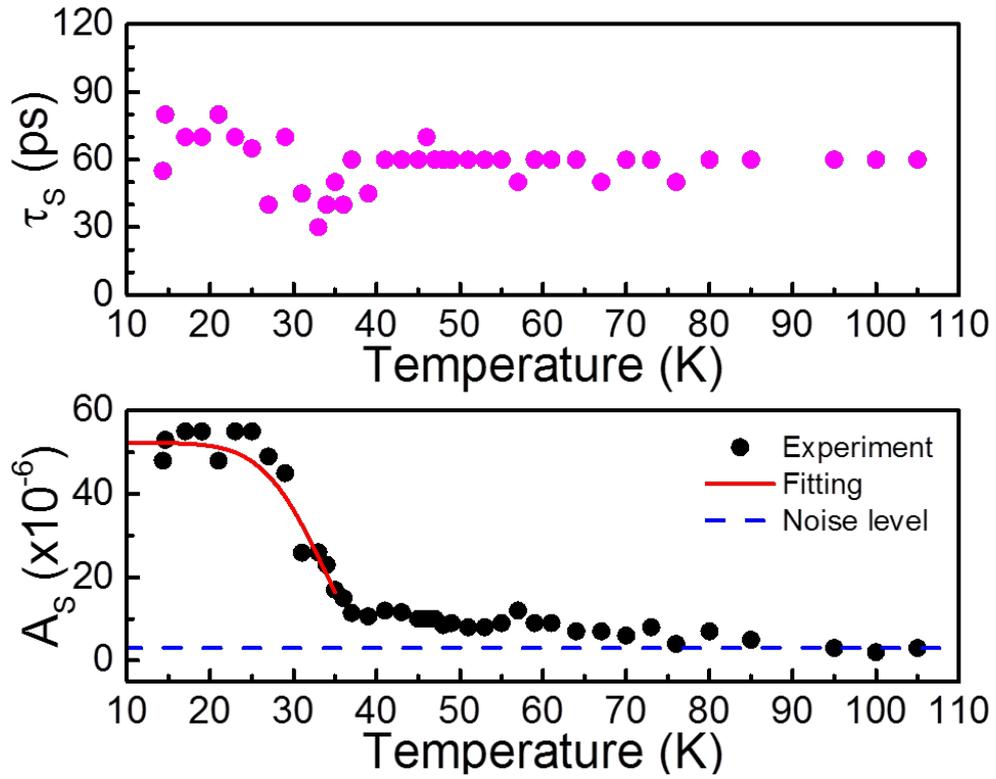

**Figure 7** The dots represent (a) $\tau_S$ and (b) $|A_S|$ of SC quasiparticle relaxation term in nearly optimally doped BKFA as a function of temperature. The red line represents the fitting curve with $\Delta(0)$=12 meV.